\documentstyle[epsfig]{mn2e}

\begin{document}

\title[Angular clustering in the SUMSS radio survey]{Angular
clustering in the SUMSS radio survey}

\author[Chris Blake et al.]{Chris Blake$^{\, 1,}$\footnotemark , Tom
Mauch$^{\, 2}$ and Elaine M. Sadler$^{\, 2}$ \\ \\ $^1$ School of
Physics, University of New South Wales, Sydney, NSW 2052, Australia \\
$^2$ School of Physics, University of Sydney, Sydney, NSW 2006,
Australia}

\maketitle

\begin{abstract}
We measure the angular correlation function of radio galaxies selected
by the 843 MHz Sydney University Molonglo Sky Survey (SUMSS).  We find
that the characteristic imprint of large-scale structure is clearly
detectable, and that the survey is very uniform.  Through comparison
with similar analyses for other wide-area radio surveys -- the 1400
MHz NRAO VLA Sky Survey (NVSS) and the 325 MHz Westerbork Northern Sky
Survey (WENSS) -- we are able to derive consistent angular clustering
parameters, including a steep slope for the clustering function,
$w(\theta) \propto \theta^{-1.1}$.  We revise upwards previous
estimates of the NVSS clustering amplitude, and find no evidence for
dependence of clustering properties on radio frequency.  It is
important to incorporate the full covariance matrix when fitting
parameters to the measured correlation function.  Once the redshift
distribution for mJy radio galaxies has been determined, these
projected clustering measurements will permit a robust description of
large-scale structure at $z \sim 0.8$, the median redshift of the
sources.
\end{abstract}

\begin{figure*}
\center
\epsfig{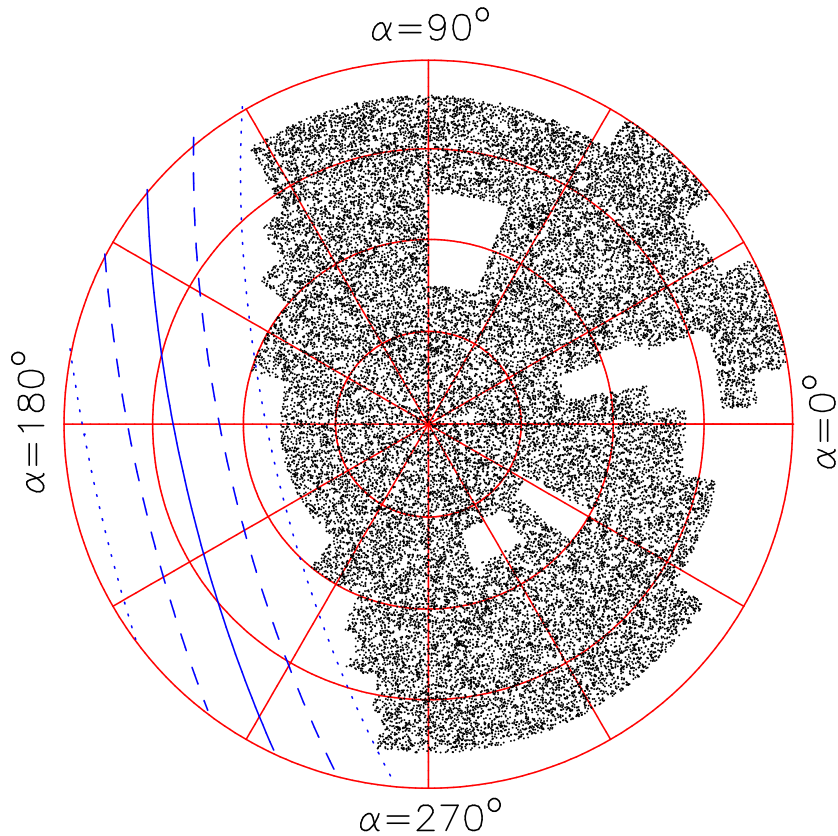}
\caption{A plot of SUMSS catalogue entries with integrated fluxes
exceeding 20 mJy in the region analyzed for large-scale structure.
The outer circumference represents declination $\delta = -50^\circ$
and the centre of the diagram is $\delta = -90^\circ$, circles being
plotted every 10 degrees.  Galactic latitudes $b = 0^\circ$, $\pm
5^\circ$ and $\pm 10^\circ$ are marked on the plot as respectively
solid, dashed and dotted lines.}
\label{figsumssplot}
\end{figure*}

\section{Introduction}
\renewcommand{\thefootnote}{\fnsymbol{footnote}}
\setcounter{footnote}{1}
\footnotetext{E-mail: chrisb@phys.unsw.edu.au}

Recent deep, wide-area radio surveys have made it possible to measure
the patterns of angular clustering in the radio sky.  Radio selection
provides a window on the galaxy distribution that differs
significantly from optical wavebands.  Radio sources probe the
high-redshift universe: the broadness of the radio luminosity function
ensures that the detected distribution of objects has a median
redshift $\overline{z} \approx 0.8$, independently of flux-density
threshold (Condon 1989).  The projected clustering signal originates
to a significant extent from redshift $\overline{z}$ and serves as a
robust tracer of structure at that epoch.  However, the wide redshift
range spanned carries a significant disadvantage: the angular
clustering is vastly diluted by the superposition of unrelated
redshift slices and hence the residual signal is faint, an order of
magnitude smaller than the projected clustering of the faintest Sloan
galaxies.  Moreover, the apparent brightness of a radio galaxy
provides no indication of its radial distance, whereas in optical
wavebands galaxies have a characteristic luminosity that may be used
to infer rough redshift information.

Radio selection is independent of Galactic and intergalactic
extinction, so that reliably complete catalogues can be obtained over
large areas.  Even so, measuring the faint imprint of clustering has
only been possible in the latest generation of surveys: Faint Images
of the Radio Sky at Twenty centimetres (FIRST, Becker, White \&
Helfand 1995), the Westerbork Northern Sky Survey (WENSS, Rengelink et
al. 1997) and the NRAO VLA Sky Survey (NVSS, Condon et al. 1998).
Each of these databases lists $> 10^5$ objects over $\sim 10,000$
deg$^2$ to limiting flux densities $S_{\rm 1400 \, MHz} \sim 3$ mJy.
The surveys differ in observing frequencies and angular resolution,
thus inter-comparisons can reveal subtle instrumental and selection
effects in individual catalogues (Blake \& Wall 2002b).

A recent addition to this complementary suite of surveys is the 843
MHz Sydney University Molonglo Sky Survey (SUMSS).  This is the first
deep wide-area radio survey mapping the southern skies; and it does so
at a frequency intermediate between that of WENSS (325 MHz) and NVSS
(1400 MHz).  Large-scale structure in the WENSS, FIRST and NVSS
surveys has been quantified by various authors: Cress et al. (1996),
Magliocchetti et al. (1998), Rengelink \& R\"ottgering (1999), Blake
\& Wall (2002a, 2002b) and Overzier et al. (2003).  In this study we
apply a first clustering analysis to the SUMSS radio survey and
compare the results with those derived from NVSS and WENSS, thus
testing whether clustering is a function of radio frequency.

It is worth emphasizing that clustering studies serve as a rigorous
test of the calibration and large-scale uniformity of a catalogue.  In
particular, the angular correlation function $w(\theta)$ analyzes the
structure present as a function of angular scale $\theta$.
Instrumental effects in a radio survey typically manifest themselves
on particular characteristic scales, and are usually rendered
transparent by such an analysis.  For example, unexpected features in
the clustering pattern within the NVSS led to corrections to the data
analysis pipeline for that survey (Blake \& Wall 2002b).

To de-project the angular clustering amplitude and recover the true
spatial clustering properties of the sample, we need to know the
redshift distribution of the sources, $N(z)$.  The radio source
population at mJy flux-density levels is a mixture of nearby
star-forming galaxies and more distant Active Galactic Nuclei (AGN)
(Condon 1989).  For example, approximately 5 per cent of NVSS sources
are detected in the 2dF Galaxy Redshift Survey (Sadler et al. 2002),
and 40 per cent of these are star-forming.  The 95 per cent of NVSS
galaxies not associated with 2dFGRS galaxies are overwhelmingly radio
AGN with redshifts $z > 0.2$, which implies that $\sim 98$ per cent of
NVSS radio sources are distant AGN.  As their optical counterparts are
often extremely faint, $N(z)$ has not yet been measured for radio AGN
at mJy flux-density levels.

However, constraints on $N(z)$ exist from luminosity-function models,
and utilizing these (with large extrapolations) we can estimate the
present-day clustering length $r_0$ of radio galaxies
(e.g. Magliocchetti et al. 1998, Blake \& Wall 2002a, Overzier et
al. 2003).  The values inferred, in the range $r_0 = 7 \rightarrow 10
\, h^{-1}$ Mpc, correspond to a clustering strength intermediate
between normal galaxies and rich clusters, consistent with the nature
of radio AGN as optically-luminous elliptical galaxies inhabiting
moderately rich environments.  An additional aim of this study is to
establish the ``best value'' of the projected clustering amplitude of
radio galaxies, to be employed in a more rigorous derivation of the
clustering length when $N(z)$ is known.

\section{The SUMSS catalogue}

\subsection{Summary}

The Sydney University Molonglo Sky Survey (SUMSS) is imaging the
southern ($\delta < -30^\circ$) radio sky at an observing frequency of
843 MHz.  The survey uses the Molonglo Observatory Synthesis Telescope
(MOST), a 1.6-km long cylindrical paraboloid reflector located near
Canberra in Australia.  The angular resolution of the SUMSS mosaics is
$45\arcsec \times$ ($45\arcsec {\rm{cosec}}{|\delta|}$).  The survey
design and science goals are described by Bock et al. (1999);
observations began in 1997 and completion is scheduled for the end of
2003.  The first instalment of the SUMSS source catalogue was
presented by Mauch et al. (2003), who showed that it is complete to
flux density $S_{\rm 843 \, MHz} = 8$ mJy for declinations $\delta <
-50^\circ$ and 18 mJy for $\delta > -50^\circ$.  SUMSS has comparable
sensitivity and resolution to the 1400 MHz NRAO VLA Sky Survey (NVSS)
in the north ($\delta > -40^\circ$).

This clustering study uses Version 1.0 of the SUMSS source catalogue,
dated 25 February 2003, available online at {\tt
www.astrop.physics.usyd.edu.au/sumsscat}.  This version of the
catalogue covers approximately 3500 deg$^2$ of the southern sky and
contains 107,765 entries.  We restricted our analysis to the region
$\delta < -50^\circ$, which has a significantly fainter completeness
limit than $\delta > -50^\circ$, and we placed additional angular
masks around the locations of the Magellanic Clouds: ($5^\circ <
\alpha < 15^\circ$, $-75^\circ < \delta < -65^\circ$) and ($70^\circ <
\alpha < 90^\circ$, $-75^\circ < \delta < -65^\circ$).  Raw SUMSS
images contain artefacts, grating rings and radial spokes (Bock et
al. 1999), typically located near very bright sources.  SUMSS
identifies these image artefacts using a reliable decision tree
algorithm; however, to be conservative, we placed circular masks of
diameter $1^\circ$ around catalogued sources brighter than 1 Jy.
After these masking procedures, 68,373 catalogue entries remained with
flux densities $S_{\rm 843 \, MHz} > 8$ mJy.  This version of the
SUMSS catalogue excludes fields located at Galactic latitudes $|b| <
10^\circ$, thus we can assume that the vast majority of the objects
are extragalactic in origin.  The sky distribution of the 20 mJy
sample is plotted in Figure \ref{figsumssplot}.

The overlap region of SUMSS and NVSS ($-40^\circ < \delta <
-30^\circ$) permits some comparisons and cross-checks.  The fraction
of SUMSS sources missing from the NVSS catalogue is $\approx 0.5$ per
cent, the majority of which are artefacts left behind by the decision
tree (these generally appear close to very bright sources which are
masked in this study as described above).  Furthermore, cross-matched
radio sources have a median spectral index $S_\nu \propto \nu^{-0.8}$
(Mauch et al. 2003 Figure 8), consistent with previous determinations
at $\nu \sim 1$ GHz.

\subsection{Uniformity of the SUMSS catalogue}
\label{secuni}

Spurious fluctuations in source surface density within a survey cause
systematic offsets in clustering measurements, unless these
fluctuations can be accurately modelled.  Large-scale density
gradients with a magnitude of $\ga 5$ per cent will drown out the
cosmological signal in the angular correlation function on scales
$\theta \ga 1^\circ$, as we discuss in Section \ref{secwenss}.  In a
wide-area survey undertaken with an interferometer such as the Very
Large Array, large-scale gradients typically appear as a function of
declination (Blake \& Wall 2002a), because the projection of the
interferometric baselines on the sky inevitably changes as a function
of declination, whereas in principle all right ascensions can be
observed at transit.  Furthermore, the ``snapshot'' radio images used
to construct a survey such as the NVSS have relatively sparse
$uv$-plane coverage and require considerable processing and cleaning
to remove artefacts.  In contrast, the $uv$-coverage of SUMSS images
is inherently continuous owing to the nature of the MOST:
deconvolution is practically unnecessary and image cleaning is
trouble-free (Bock et al. 1999).

\begin{figure}
\center
\epsfig{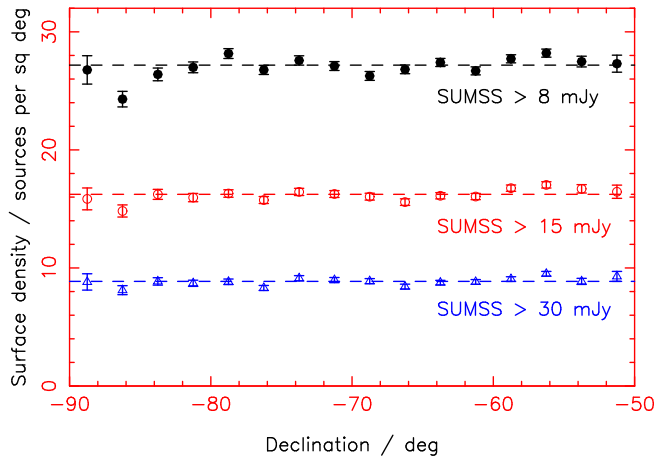}
\caption{SUMSS source surface density as a function of declination.
Sources are binned in declination bands for flux-density thresholds 8
mJy (solid circles), 15 mJy (open circles) and 30 mJy (triangles).
The dashed lines indicate the average surface density at each
threshold.  The error in the number of sources $N$ in a band is
plotted as the `'Poisson error'' $\sqrt{N}$.  Masked areas are
excluded from this analysis.}
\label{figdenssumss}
\end{figure}

Figure \ref{figdenssumss} plots the source surface density of the
SUMSS catalogue as a function of declination for flux-density
thresholds 8 mJy, 15 mJy and 30 mJy.  There are no apparent spurious
gradients as a function of declination, although the scatter of the
data points exceeds that predicted by Poisson statistics (the error
bars).  This excess variance is produced by clustering, by
multiple-component sources and (potentially) by spurious fluctuations
in surface density.  We placed a quantitative upper limit on the rms
spurious density fluctuations by assuming that these were entirely
responsible for the excess variance.  We modelled the surface density
at declination $\delta$ as a sum of two components, $\sigma(\delta) =
\sigma_0 + \sigma_1(\delta)$, where $\sigma_0$ is a constant-density
``unperturbed'' level and $\sigma_1(\delta)$ is a fluctuating offset
(such that $\overline{\sigma_1} = 0$).  The variance of the measured
density over the declination bins is the sum of a ``Poisson''
contribution (due to $\sigma_0$) and a ``systematic'' contribution
(due to $\sigma_1$).  We derived the variance of $\sigma_1$ by
subtracting the contribution of Poisson statistics from the total
variance; the fractional rms fluctuation is then $f_{\rm fluc} =
\sqrt{{\rm Var}(\sigma_1)}/\sigma_0$.  The resulting upper limits on
spurious density fluctuations in the SUMSS catalogue are respectively
$2.2\%$, $2.0\%$ and $2.5\%$ at thresholds 8 mJy, 15 mJy and 30 mJy.

A systematic variation in surface density $\sigma$ creates an
approximately constant offset in $w(\theta)$ of magnitude
$\overline{\epsilon^2}$, where $\epsilon =
(\sigma-\overline{\sigma})/\overline{\sigma}$ is the surface
overdensity (Blake 2002).  Thus we can write:
\begin{equation}
\Delta w(\theta) = \overline{\epsilon^2} = \frac{{\rm
Var}(\sigma_1)}{\sigma_0^2} = (f_{\rm fluc})^2
\end{equation}
Our measurements detailed in the previous paragraph therefore placed
an upper limit $\Delta w(\theta) \approx (0.02)^2 = 4 \times 10^{-4}$
on the effect of source density gradients in SUMSS.  Thus our
determination of $w(\theta)$ in Section \ref{seccorrsumss} is not
significantly affected by any such gradients.

\section{Angular correlation function for the SUMSS catalogue}
\label{seccorrsumss}

We quantified the projected clustering in the SUMSS catalogue using
the angular correlation function, $w(\theta)$.  This two-point
statistic is determined by counting the number of pairs of catalogue
entries, $DD(\theta)$, as a function of angular separation $\theta$,
and comparing the result to the number of pairs expected by random
chance.  The value of $w(\theta)$ is the fractional enhancement of
pair counts above random.  We computed $w(\theta)$ using the
Landy-Szalay estimator (Landy \& Szalay 1993).  Using the Hamilton
estimator (Hamilton 1993) led to identical results.  The error in each
separation bin is plotted in the Figures as a ``Poisson'' error:
\begin{equation}
\Delta w(\theta) = \frac{1 + w(\theta)}{\sqrt{DD(\theta)}}
\label{eqwerr}
\end{equation}
This is known to be a very good approximation for the variance of this
estimator of the correlation function.

When fitting models to the measured $w(\theta)$, we used the full
covariance matrix of the separation bins, $C_w$.  The covariance
matrix for a given correlation function model was derived using the
analytic approximation of Bernstein (1994, equation 39), who presents
a specific treatment of the Landy-Szalay estimator; see also
Eisenstein \& Zaldarriaga (2001).  Bernstein's approximations -- $N
\gg 1$, $w(\theta) \ll 1$, $w_\Omega \ll w(\theta)$ -- are valid for
the present surveys.  We made an additional approximation by
neglecting higher-order correlations ($q = 0$).  Once the correlation
matrix has been computed, the chi-squared statistic is given by
\begin{equation}
\chi^2 = {\bf \Delta w}^T C_w^{-1} {\bf \Delta w}
\end{equation}
where ${\bf \Delta w}$ is a vector containing the differences between
the measured values of $w(\theta)$ and the model.

Apparent clustering in radio source catalogues is due to two effects:
the cosmological clustering of separate galaxies, and the resolution
of individual radio galaxies into multiple components which appear as
separate catalogue entries.  A $w(\theta)$ analysis neatly separates
these two contributions.  Excess pairs with separations $\theta \la
0.1^\circ$ are almost exclusively produced by multiple components;
cosmological effects pre-dominate on scales $\theta \ga 0.1^\circ$.  A
good fitting formula for the $w(\theta)$ data is a sum of two
power-laws with sharply differing slopes, resulting in a clear break
at $\theta \approx 0.1^\circ$ (see Blake \& Wall 2002a and Overzier et
al. 2003 for further justification of this model).

\begin{figure}
\center
\epsfig{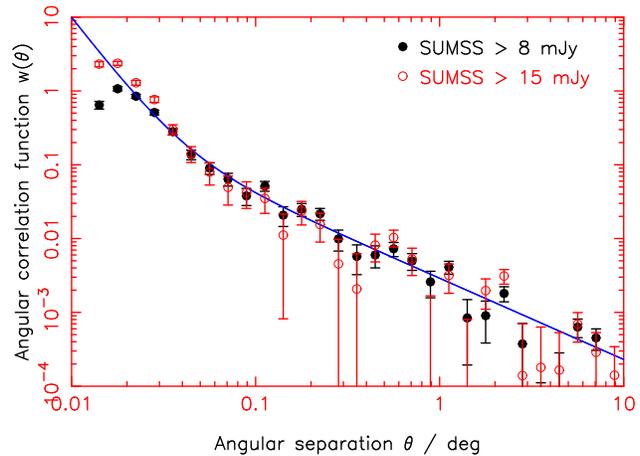}
\caption{Measurement of the SUMSS angular correlation function at
flux-density thresholds $S_{\rm 843 \, MHz} = 8$ mJy (solid circles)
and 15 mJy (open circles).  The best-fitting sum of two power-laws for
the 8 mJy data is overplotted.  This is not a good fit, and we
preferred to use a more conservative threshold of 10 mJy for the
subsequent analysis.}
\label{figcorrsumss}
\end{figure}

These effects are, as anticipated, present in the SUMSS catalogue.
Figure \ref{figcorrsumss} plots measurements of the SUMSS $w(\theta)$
at flux-density thresholds 8 mJy (which is the completeness limit) and
15 mJy.  At 8 mJy, a sum of two power-laws is not in fact a good fit
to the data points (for $\theta > 0.02^\circ$), with a minimum reduced
chi-squared statistic $\chi^2_{\rm red} = 2.16$ (derived using the
full covariance matrix; the probability that $\chi^2$ should exceed
this value by chance is $Q = 1.0 \times 10^{-3}$).  However, at 10 mJy
the fit has improved to $\chi^2_{\rm red} = 1.66$ ($Q = 0.024$) and
the corresponding value for 20 mJy is $\chi^2_{\rm red} = 1.37$ ($Q =
0.11$).  The poor fit at 8 mJy is probably due to unidentified
systematic effects near the completeness limit of the survey, and we
used the more conservative threshold of 10 mJy for subsequent
analysis (see Section \ref{secampcomp}).

We investigated the dependence on flux density of the two power-laws
used to model $w(\theta)$.  As suggested by Figure \ref{figcorrsumss}
and quantified in Section \ref{secampflx}, we found no change with
flux density in the amplitude of the shallower power-law, which
dominates on scales $\theta \ga 0.1^\circ$ and reflects cosmological
clustering.  This finding is in agreement with other wide-area radio
surveys in the flux density range $3$ mJy $< S_{\rm 1400 \, MHz} < 50$
mJy.  Note that this result is in sharp contrast to clustering
measured in optical wavebands, where the clustering amplitude declines
rapidly as galaxies become fainter.  The explanation for this
difference is simple: unlike in optical wavebands, fainter radio
galaxies are not (on average) more distant.  As shown in Figure
\ref{figcorrsumss}, the multiple-components piece of $w(\theta)$ (at
small separations) has an amplitude that decreases with flux density.
Blake \& Wall (2002a) demonstrated that the expected dependence of
this amplitude is $1/\sigma$, where $\sigma$ is the source surface
density.

At very small separations $\theta < 0.02^\circ$ there is a fall-off in
the value of $w(\theta)$ with decreasing $\theta$.  This is not real,
but is a signal-to-noise effect caused by the failure of the survey to
resolve weak double sources with separations slightly greater than the
beam-width.  At very large separations $\theta \approx 10^\circ$, the
value of $w(\theta)$ is consistent with zero.  This is strong evidence
for a high degree of uniformity in the survey: as detailed in Section
\ref{secuni}, calibration gradients produce spurious offsets in the
value of $w(\theta)$.

\section{Comparison with other wide-area radio surveys}

In this Section we compare the SUMSS clustering results with
equivalent analyses of the 1400 MHz NRAO VLA Sky Survey (NVSS, Condon
et al. 1998) and the 325 MHz Westerbork Northern Sky Survey (WENSS,
Rengelink et al. 1997).  These different samples are summarized in
Table \ref{tabsamp}.  The three surveys have comparable angular
resolutions but different observing frequencies, potentially probing
different selections of the radio source population.  It is of
interest to discover whether clustering is a function of radio
frequency.  Cosmological clustering of radio galaxies is independent
of flux density in the range $3$ mJy $< S_{\rm 1400 \, MHz} < 50$ mJy
(Section \ref{secampflx}), thus the choice of comparison flux-density
threshold for NVSS and WENSS is not important.

\begin{table*}
\center
\caption{Summary of radio source samples used in this investigation.}
\label{tabsamp}
\begin{tabular}{cccccccc}
\hline Survey & Frequency & Angular resolution & Declination & Angular
area & Flux density & Number of & Surface density \\ & (MHz) &
(arcsec) & range & (deg$^2$) & (mJy) & sources & (deg$^{-2}$) \\
\hline

SUMSS & 843 & $45 \times (45 \, {\rm{cosec}} |\delta|)$ & $-90^\circ <
\delta < -50^\circ$ & 2,516 & $> 8$ & 68,373 & 27.2 \\

& & & & & $> 10$ & 57,071 & 22.7 \\

& & & & & $> 15$ & 40,843 & 16.2 \\

& & & & & $> 30$ & 22,301 & 8.9 \\

NVSS & 1400 & 45 & $-40^\circ < \delta < 90^\circ$ & 31,104 & $> 10$ &
524,808 & 16.9 \\

WENSS & 325 & $54 \times (54 \, {\rm{cosec}} |\delta|)$ & $30^\circ <
\delta < 75^\circ$ & 8,409 & $> 35$ (int) & 107,937 & 12.8 \\

& & & $30^\circ < \delta < 60^\circ$ & 6,579 & $> 35$ (peak) & 87,501
& 13.3 \\

\hline
\end{tabular}
\end{table*}

\subsection{NVSS}

\begin{figure}
\center
\epsfig{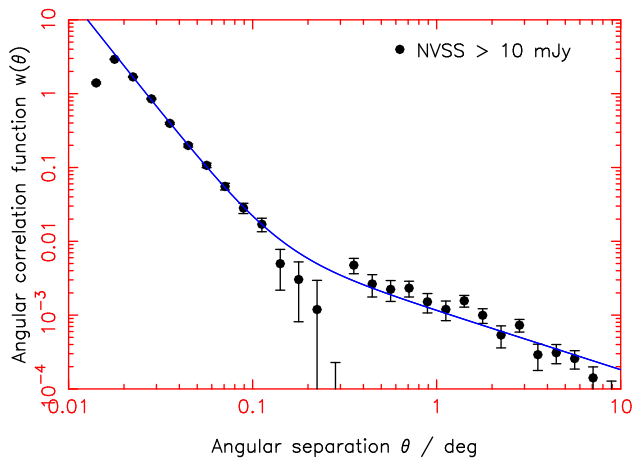}
\caption{Measurement of the NVSS angular correlation function at
flux-density threshold $S_{\rm 1400 \, MHz} = 10$ mJy, from Blake \&
Wall (2002a).  The best-fitting sum of two power-laws is overplotted.
This is not a good fit owing to a deficit of pairs with separations
$\theta \approx 0.2^\circ$.}
\label{figcorrnvss}
\end{figure}

The angular correlation function for NVSS has been measured by Blake
\& Wall (2002a) and Overzier et al. (2003).  The result of Blake \&
Wall for flux-density threshold $S_{\rm 1400 \, MHz} = 10$ mJy is
plotted in Figure \ref{figcorrnvss}; this is the deepest flux density
for which systematic surface density gradients are negligible.  A sum
of two power-laws is an excellent fit to the data points, except for
separations $0.1^\circ < \theta < 0.3^\circ$, for which there is a
significant deficit of pairs.  This results in the overall best fit
being poor ($\chi^2_{\rm red} = 2.21$, $Q = 7.0 \times 10^{-4}$).

We made the following improvement to previous analyses of the NVSS
$w(\theta)$: we suggest that the data points at $\theta \approx
0.2^\circ$ are suspect.  The NVSS synthesized beam features very
strong, broad diffraction spikes (sidelobes) offset by $0.2^\circ
\rightarrow 0.3^\circ$ from the main lobe, with amplitudes which are
almost 30 per cent of the peak (Condon et al. 1998 Section 4.1 and
Figure 17).  These diffraction features correspond to the shortest
baselines used in the VLA observations, which are about 37 metres
($\approx 170$ wavelengths).

It is very plausible that the required deep cleaning of the NVSS
images causes the small deficit of galaxy pairs with separations
$\theta \approx 0.2^\circ$.  The consequent CLEAN bias will not remove
10 mJy sources near stronger sources, but it might reduce the flux
density of the fainter source in each $0.2^\circ$ pair by some
fraction of the rms noise, which is $\sigma_{\rm noise} = 0.5$ mJy,
independent of the true source flux density (J. Condon, priv. comm.).
The observed shortfall of $0.2^\circ$ pairs is $\approx 1$ per cent
(Figure \ref{figcorrnvss}).  In comparison, the number of data pairs
in the separation range $0.1^\circ < \theta < 0.3^\circ$ is increased
by 1.6 per cent when the flux-density threshold is reduced by
$\sigma_{\rm noise}/5$ from 10.0 mJy to 9.9 mJy; we conclude that the
shortfall of $0.2^\circ$ pairs is probably an instrumental effect.

The effect of the deficit of pairs with separations $\theta \approx
0.2^\circ$ is to reduce spuriously the amplitude of the fitted
clustering power-law, which must attempt to accommodate these low
points.  Fitting the clustering power-law to the separation range
$\theta = 0.3^\circ \rightarrow 10^\circ$ instead, a much better fit
to the data was obtained ($\chi^2_{\rm red} = 1.07$, $Q = 0.38$) with
a clustering amplitude $\sim 50$ per cent higher than previously
recorded.  We postpone further discussion of the best-fitting
clustering parameters to Section \ref{secampcomp}.

Overzier et al. (2003) describe an ``unexplained bump'' at separations
$0.1^\circ < \theta < 0.3^\circ$ in the NVSS $w(\theta)$ measured
above a threshold of 200 mJy.  This is very probably an additional
consequence of the diffraction sidelobes.

\subsection{WENSS}
\label{secwenss}

The first measurements of the WENSS angular correlation function were
presented by Rengelink \& R\"ottgering (1999); we performed our own
investigation.  We analyzed the ``main'' WENSS catalogue, which covers
the region $0^\circ < \alpha < 360^\circ$, $30^\circ < \delta <
75^\circ$, and is available online at {\tt
www.strw.leidenuniv.nl/$\sim$dpf/wenss}.  The catalogue lists merged
multiple-component (``M'') sources as well as their separate
components (``C'' sources); we omitted the ``M'' sources.  In addition
we masked areas within $5^\circ$ of the Galactic plane, and inside a
large box ($15^\circ \times 15^\circ$) surrounding the bright radio
source Cygnus A.

\begin{figure}
\center
\epsfig{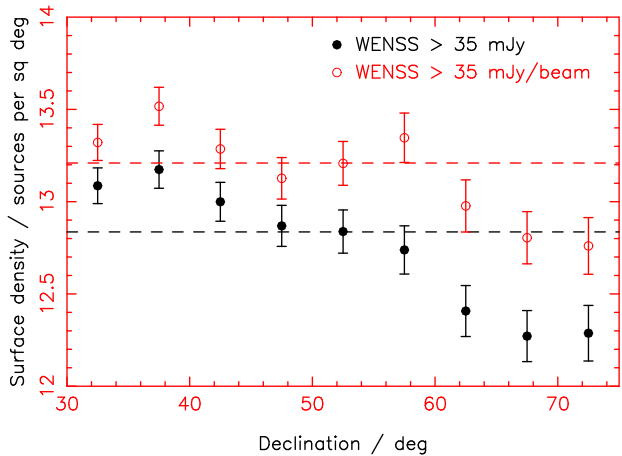}
\caption{WENSS source surface density as a function of declination at
integrated flux-density threshold 35 mJy (solid circles) and peak
flux-density threshold 35 mJy/beam (open circles).  The dashed lines
indicate the average surface densities at these thresholds.  The error
in the number of sources $N$ in a band is $\sqrt{N}$.}
\label{figdenswenss}
\end{figure}

The completeness limit of WENSS has been determined as $S_{\rm 325 \,
MHz} = 35$ mJy (Rengelink \& R\"ottgering 1999).  Figure
\ref{figdenswenss} plots the WENSS catalogue surface density at this
limit as a function of declination in two different ways: firstly for
a 35 mJy threshold in (deconvolved) integrated flux density, and
secondly for a 35 mJy/beam threshold in (convolved) peak flux density.
For a point source the integrated and peak flux densities are equal;
the large WENSS beam ensures that these quantities do not usually
differ significantly for catalogue entries.

\begin{figure}
\center
\epsfig{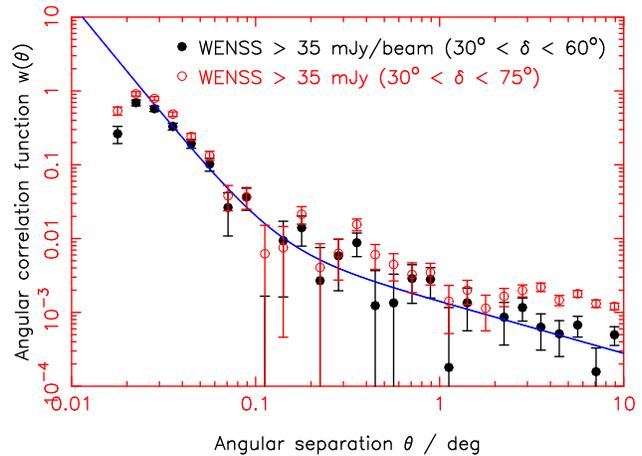}
\caption{Measurement of the WENSS angular correlation function. The
open circles plot the result for integrated flux-density threshold
$S_{\rm 325 MHz} = 35$ mJy over declination range $30^\circ < \delta <
75^\circ$.  A spurious offset is evident for separations $\theta \ga
1^\circ$ due to a source surface density gradient.  This motivated us
to select a more uniform catalogue by imposing a 35 mJy/beam threshold
in peak flux density over the area bounded by $30^\circ < \delta <
60^\circ$.  The $w(\theta)$ result for this second selection is
plotted as the solid circles, together with the best-fitting double
power-law model.}
\label{figcorrwenss}
\end{figure}

For the threshold in integrated flux density in Figure
\ref{figdenswenss}, there is a systematic gradient amounting to a
$\sim 10$ per cent fluctuation between $\delta = 30^\circ$ and $\delta
= 75^\circ$.  A possible explanation for this feature is that at
higher declinations, the projection of the interferometric baselines
on the sky is shorter, implying reduced surface brightness sensitivity
and marginally reduced completeness as regards integrated flux.  This
surface density gradient creates a significant spurious offset in
$w(\theta)$, drowning out the cosmological signal at separations
$\theta \ga 1^\circ$ (see Figure \ref{figcorrwenss}).  These
systematic effects are much less serious for the sample selected by
peak flux density.  Thus, motivated by Figure \ref{figdenswenss}, we
created a more uniform WENSS source catalogue by imposing a peak
flux-density threshold of 35 mJy/beam, and restricting our analysis to
the declination band $30^\circ < \delta < 60^\circ$ ($\approx 80$ per
cent of the available area).

Our measurements of the WENSS $w(\theta)$ for these two catalogue
selections are displayed in Figure \ref{figcorrwenss}.  For the peak
flux-density threshold, a double power-law again provides a reasonable
fit ($\chi^2_{\rm red} = 1.50$, $Q = 6.4 \times 10^{-2}$).  However,
the influence of surface gradients is still not negligible and the
angular correlation function at large separations ($\theta \sim
10^\circ$) is not consistent with zero.  Therefore the best-fitting
clustering power-law spuriously acquires a significantly shallower
slope than that determined for SUMSS and NVSS.  We therefore added an
additional parameter -- a constant offset -- to the WENSS angular
correlation function fitting formula.  Although this extra parameter
does not significantly improve the $\chi^2$ statistic of the
best-fitting model, it is well-motivated by the existence of the
surface density gradients.

\subsection{Comparison of clustering amplitudes}
\label{secampcomp}

\begin{figure}
\center
\epsfig{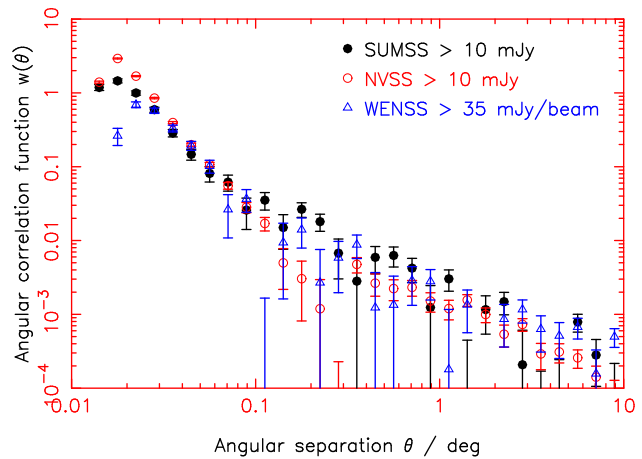}
\caption{Overplot of the angular correlation functions for SUMSS ($>
10$ mJy, solid circles), NVSS ($> 10$ mJy, open circles) and WENSS ($>
35$ mJy/beam, triangles).}
\label{figcorrall}
\end{figure}

We compared the angular correlation functions for SUMSS ($> 10$ mJy),
NVSS ($> 10$ mJy) and WENSS ($> 35$ mJy/beam) (Figure
\ref{figcorrall}).  We used the 10 mJy threshold for SUMSS (rather
than the completeness limit of 8 mJy) because 10 mJy is the deepest
flux-density level for which a power-law model proved a good fit to
the data.

\begin{figure}
\center
\epsfig{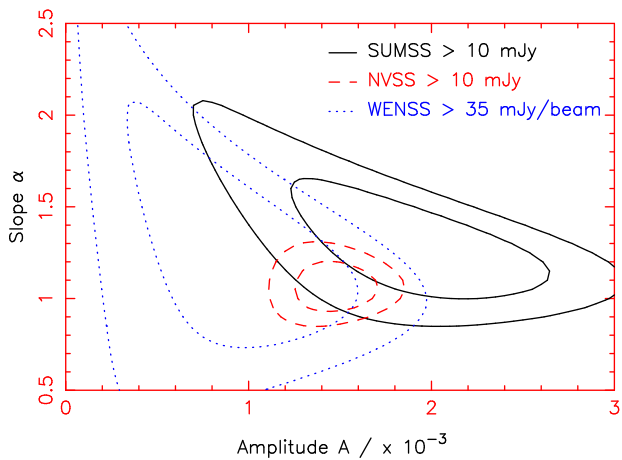}
\caption{Measurement of the clustering parameters $w(\theta) = A \,
\theta^{-\alpha}$ for SUMSS (solid line), NVSS (dashed line) and WENSS
(dotted line).  Contours of constant $\chi^2$ are shown in the space
of the parameters ($A$, $\alpha$), corresponding to $1\sigma$ and
$2\sigma$ confidence regions.  As the plot is in the space of two
varying parameters, the $1\sigma$ and $2\sigma$ contours are defined
by $\chi^2$ increasing by respectively 2.30 and 6.17 from its minimum.
When generating this plot, the covariance matrix was held fixed at
that corresponding to the best-fitting correlation function model.}
\label{figcorrparam}
\end{figure}

We fitted a power-law $w(\theta) = A \, \theta^{-\alpha}$ to the
cosmological clustering part of each correlation function, using the
full covariance matrix.  In the case of WENSS, we introduced a third
variable parameter, a constant offset, to accommodate the surface
gradients.  The angular ranges to which the functions were fitted were
$\theta > 0.2^\circ$ (SUMSS, WENSS) and $\theta > 0.3^\circ$ (NVSS).
For SUMSS and WENSS, $\theta > 0.2^\circ$ safely exceeds separations
for which multiple component sources are important ($\theta \la
0.1^\circ$).  For NVSS, $\theta > 0.3^\circ$ also avoids the influence
of diffraction spikes at $\theta \approx 0.2^\circ$.  Figure
\ref{figcorrparam} plots contours of constant $\chi^2$ in the space of
the clustering parameters ($A$, $\alpha$).  The results for the three
independent surveys are consistent; there is no evidence that
clustering is a function of observing frequency.

\begin{table*}
\center
\caption{Results of power-law fits to the cosmological clustering part
of the angular correlation function, $w(\theta) = A \,
\theta^{-\alpha}$, for SUMSS, NVSS and WENSS.  In the case of WENSS a
constant offset is also fitted to the function.  The best fit is
obtained by minimizing the $\chi^2$ statistic, incorporating the full
covariance matrix.  The errors in the parameters are derived by
varying each in turn from the best-fitting combination (keeping the
others fixed) and determining the variation for which $\Delta \chi^2 =
1$, the appropriate 1-sigma increment when varying one fitted
parameter.}

\label{tabcorr}
\begin{tabular}{ccccccc}
\hline Survey & Flux density & Amplitude $A$ & Slope $\alpha$ &
Constant offset & $\chi^2_{\rm red}$ & $Q$ \\ & (mJy) & ($\times
10^{-3}$) & & ($\times 10^{-3}$) & & \\ \hline

SUMSS &  $> 10$ & $2.04  \pm 0.38$ & $1.24  \pm 0.16$ & --  & $1.79$ &
$0.03$ \\

NVSS & $> 10$ & $1.49 \pm 0.15$ & $1.05 \pm 0.10$ & -- & $0.89$ &
$0.57$ \\

WENSS & $> 35$ (peak) & $1.01 \pm 0.35$ & $1.22 \pm 0.33$ & $0.35 \pm
0.19$ & $1.37$ & $0.15$ \\

\hline
\end{tabular}
\end{table*}

The best-fitting clustering parameters for the three surveys are
tabulated in Table \ref{tabcorr}.  They differ somewhat from those
previously reported for the NVSS, which were $A = (1.0 \pm 0.2) \times
10^{-3}$, $\alpha = 0.8 \pm 0.1$ (Blake \& Wall 2002a, Overzier et
al. 2003; note that $\theta$ is measured in degrees for all quoted
values of the amplitude $A$).  These differences are due to ignoring
the low $w(\theta)$ data at $\theta \approx 0.2^\circ$ in the NVSS
measurement.  The best-fitting slope is now somewhat higher, $\alpha
\approx 1.1$, agreeing with the value originally reported for the
FIRST survey by Cress et al. (1996).  In fact, it is not surprising
that the clustering slope should exceed the canonical value $\alpha =
0.8$ of normal galaxies; it is known that $\alpha$ depends on galaxy
type and takes on its steepest value for elliptical galaxies such as
radio AGN (Norberg et al. 2002).  The new value of the clustering
amplitude, $A \approx 1.6 \times 10^{-3}$, also exceeds the previous
determination.

Inclusion of the full covariance matrix in the fitting procedure
increases the errors on the clustering parameters by $\sim 50$ per
cent, compared to the approximation of neglecting correlations between
separation bins.  This result holds despite the fact that the
variances of each bin are dominated by Poisson noise rather than by
cosmic variance.

The multiple-components piece of $w(\theta)$, at $\theta \la
0.1^\circ$, varies between the three surveys (Figure
\ref{figcorrall}).  These differences are explained by the $1/\sigma$
dependence of the amplitude of the small-angle $w(\theta)$ on surface
density $\sigma$, together with the small variations in angular
resolution between the surveys.

\subsection{Projected clustering amplitude as a function of flux density}
\label{secampflx}

\begin{figure}
\center
\epsfig{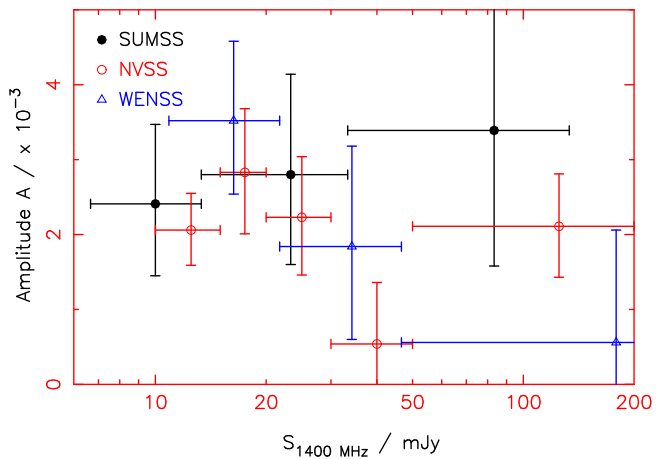}
\caption{Angular correlation function amplitudes for SUMSS, NVSS and
WENSS for various independent flux-density ranges.  The slope of
$w(\theta)$ was fixed at $-1.1$ in all cases.}
\label{figflxall}
\end{figure}

We repeated the angular correlation function measurements for SUMSS,
NVSS and WENSS for various independent flux-density ranges, and fitted
a power-law $w(\theta) = A \, \theta^{-1.1}$ to the cosmological
clustering parts of each function (keeping the slope fixed to permit
comparison of amplitudes).  We fitted an additional constant offset in
the case of the WENSS analysis.  The amplitudes are displayed in
Figure \ref{figflxall}; results at different observing frequencies
have been transformed to 1400 MHz assuming a spectral index $S_\nu
\propto \nu^{-0.8}$.

Figure \ref{figflxall} demonstrates that the amplitude of projected
clustering does not depend on either observing frequency or flux
density (for $S_{\rm 1400 \, MHz} \la 50$ mJy).  At brighter flux
densities the signal-to-noise of the measurement is too low to allow
firm conclusions to be drawn.  We note that Overzier et al. (2003)
found evidence for an enhanced angular clustering amplitude in NVSS
for a threshold $S_{\rm 1400 \, MHz} = 200$ mJy, with a large error
bar.

\section{Summary}

The results of this investigation may be summarized as follows:
\begin{itemize}
\item The angular correlation function for the SUMSS radio survey is
measured for the first time and is found to have the same double
power-law structure as previously identified in NVSS.
\item Our analysis reveals no evidence for large-scale systematic
gradients in SUMSS, such as might be imprinted by instrumental effects
or data reduction.
\item The NVSS $w(\theta)$ is re-examined, and data points on angular
scales $0.1^\circ < \theta < 0.3^\circ$ are excluded from the fitting
process due to the possible influence of imperfectly-cleaned bright
sidelobes.
\item The WENSS $w(\theta)$ is re-examined, and a systematic surface
density gradient as a function of declination is identified for
sources selected above a fixed threshold in integrated flux density.
A more uniform source catalogue can be created by imposing a threshold
in peak flux density.
\item The parameters of the clustering power-law, $w(\theta) = A \,
\theta^{-\alpha}$, derived for SUMSS, NVSS and WENSS, are found to be
in agreement, having values $\alpha \approx 1.1$ and $A \approx 1.6
\times 10^{-3}$.  Errors in the parameters for the individual surveys
are mapped in Figure \ref{figcorrparam} and Table \ref{tabcorr}.
\item Inclusion of the full covariance matrix of the separation bins
significantly increases the errors on the fitted clustering
parameters, compared to the assumption that the bins are statistically
independent.
\item We find no evidence for a dependence of projected clustering
amplitude on either flux density or observing frequency.
\end{itemize}
The inter-comparison of these three wide-area radio surveys has
resulted in the refinement of our knowledge of the projected
clustering of radio galaxies.  Once the redshift distribution of these
sources has been determined at mJy flux-density levels, their angular
clustering amplitude may be used to make a robust measurement of the
spatial clustering of luminous elliptical galaxies at $z \approx 0.8$.

\section*{Acknowledgments}

We acknowledge useful discussions with Carole Jackson concerning radio
source populations as a function of frequency, with Jim Condon
concerning instrumental effects in the NVSS, and with Michael Brown
concerning the covariance of correlation function estimates.  We thank
Jasper Wall, Roderik Overzier and Dick Hunstead for useful comments on
earlier drafts of the manuscript.  We acknowledge Roeland Rengelink
for a very helpful referee's report.


\begin{thebibliography}{}
\bibitem{1} Becker R.H., White R.L., Helfand D.J., 1995, ApJ, 450, 559
\bibitem{2} Bernstein G.M., 1994, ApJ, 424, 569
\bibitem{3} Blake C.A., 2002, PhD thesis, University of Oxford
\bibitem{4} Blake C.A., Wall J.V., 2002a, MNRAS, 329, L37
\bibitem{5} Blake C.A., Wall J.V., 2002b, MNRAS, 337, 993
\bibitem{6} Bock D.C.-J., Large M.I., Sadler E.M., 1999, ApJ, 117, 1578
\bibitem{7} Condon J.J., 1989, ApJ, 338, 13
\bibitem{8} Condon J.J., Cotton W.D., Greisen E.W., Yin Q.F., Perley
R.A., Taylor G.B., Broderick J.J., 1998, AJ, 115, 1693
\bibitem{9} Cress C.M., Helfand D.J., Becker R.H., Gregg M.D., White
R.L., 1996, ApJ, 473, 7
\bibitem{10} Eisenstein D.J., Zaldarriaga M., 2001, ApJ, 546, 2
\bibitem{11} Hamilton A.J.S., 1993, ApJ, 417, 19
\bibitem{12} Landy S.D., Szalay A.S., 1993, ApJ, 412, 64
\bibitem{13} Magliocchetti M., Maddox S., Lahav O., Wall J., 1998,
MNRAS, 300, 257
\bibitem{14} Mauch T., Murphy T., Buttery H.J., Curran J., Hunstead
R.W., Piestrzynski B., Robertson J.G., Sadler E.M., 2003, MNRAS, 342,
1117
\bibitem{15} Norberg et al., 2002, MNRAS, 332, 827
\bibitem{16} Overzier R.A., R\"ottgering H.J.A., Rengelink R.B.,
Wilman R.J., 2003, A\&A, 405, 53
\bibitem{17} Rengelink R.B., Tang Y., de Bruyn A.G., Miley G.K.,
Bremer M.N., R\"ottgering H.J.A., Bremer M.A.R., 1997, A\&AS, 124, 259
\bibitem{18} Rengelink R.B., R\"ottgering H.J.A., 1999, in ``The Most
Distant Radio Galaxies'', ed. R\"ottgering H.J.A., Best P.N., Lehnert
M.D., p.399
\bibitem{19} Sadler E.M. et al., 2002, MNRAS, 329, 227
\end{thebibliography}
\end{document}